\documentstyle[prd,aps,preprint]{revtex}
\begin{document}

\draft

%
%  Uncomment following two lines and one below for 2 column format.
%@twocolumnfalse\endcsname

\preprint{Nisho-99/4} \title{Emission of Radio Waves in Gamma Ray Bursts\\
and Axionic Boson Stars} 
\author{Aiichi Iwazaki}
\address{Department of Physics, Nishogakusha University, Chiba
  277-8585,\ Japan.} \date{August 15, 1999} \maketitle
\begin{abstract}
We point out that the bursts of photons 
with the energy of the axion mass may appear coincidentally 
with gamma ray bursts 
if the gamma ray bursts are caused by collisions between neutron stars and 
axionic boson stars. In this mechanism, jets are formed 
in the collisions with large Lorentz factors $\geq 10^2$.
We explain qualitatively time-dependent complex structures of gamma ray bursts 
as well as the large energy problem.
Therefore, with detection of the monochromatic photons 
we can test the model and determine the axion mass.
\end{abstract}
%\pacs{73.61.-r,73.20.Dx,73.40.Hm,73.40.Gk}

\vskip .3cm

\pacs{14.80.Mz, 98.80.Cq, 97.60.Jd, 98.70.Rz, 95.30.+d, 05.30.Jp, 
98.70.-f \\Axion, Gamma Ray Burst, Dark Matter, Boson Star} 
%\hspace*{3cm}}
\vskip2pc
%%%%%%%%%%%%%%%%%%%%%%%%%%%%%%%%%%%%%%%%%%%%%%%%
\tightenlines
One of the most fascinating problems in astrophysics is the mechanism
causing gamma ray bursts ( GRBs ). 
Up to now there is no satisfactory mechanism explaining GRB 
with time-dependent complex structures although plausible models\cite{model}
have been proposed. On the other hand, one of the most fascinating problems 
in cosmology is the dark matter in the Universe\cite{text}. 
Axion\cite{PQ,kim} is one of most plausible candidates for the dark matter 
although 
even its existence has not yet been confirmed. 
It is important to find the relic\cite{iw} of the axion 
or the axion itself\cite{s}.
Probably, some of axions may form boson stars in the present Universe
by gravitational cooling\cite{cooling} or 
gravitational collapse of axion clumps formed at the period of 
QCD phase transition\cite{kolb}.

In this letter we analyze a mechanism of generating GRBs 
proposed by the author\cite{iwa},
which is a collision between 
an axionic boson star\cite{real,iwa1} and a neutron star.
The axionic boson star dissipates its energy very rapidly inside of
the neutron star, which leads to jets formed by baryons emitted
in the collision. 
This dissipation\cite{iwa1} of the axion star energy is the engine 
for GRB in our model.
The energy released reaches as much as  $10^{-4}\,M_{\odot}$.
%We show that the bursts of monochromatic electromagnetic radiations with 
%the frequency 
%( $2.4\times 10^9\,m_5\,\mbox{Hz}$ with $m_5=m_a/10^{-5}\,\mbox{eV}$ )
%given by the axion mass $m_a$ appear coincidentally with GRBs.
Furthermore, in the collision, heavy nuclei or protons composing the outer
crust of a neutron star are accelerated to obtain energies 
larger than $10^2\,\mbox{GeV}$. We argue that these particles may form jets 
with Lorentz factors larger than $10^2$ and with small solid angles 
$\sim 10^{-4}$. Therefore, 
the amount of the energy released in the collision is consistent with
the observations.
In general the collisions may occur several times when the axion star 
trapped to the neutron star falls on to the neutron star; the axion star 
collides with the neutron star without losing its whole energy
and passes through the neutron star, and then goes back
and collides again with the neutron star. This process continues 
until the whole energy of axion star is lost.
Each one of the collisions produces 
a fireball and consequently they generate time-dependent complex 
structures\cite{complex} of GRB.

There have been several other models\cite{model} proposed for the mechanisms 
of generating GRBs: Merger of compact stars, hypernovae or supranovae.
Observational clear distinction of these models seems to be difficult
until further observations of GRBs and more detail analyses of them 
are performed. However, in our model 
we can predict a clear event associated with
a GRB: appearance of monochromatic electromagnetic radiations with 
the frequency 
( $2.4\times 10^9\,m_5\,\mbox{Hz}$ with $m_5=m_a/10^{-5}\,\mbox{eV}$ )
given by the axion mass $m_a$. The bursts may appear just before 
or coincidentally with the GRB. Therefore, the model 
can be easily tested observationally. If the bursts are 
observed, we can not only confirm the existence of axion but also detect its
mass.

%the bursts of radio waves with 
%the Compton wave length of an axion ( $13\,m_5^{-1}\,\mbox{cm}$ ). 
%The bursts may appear just before 
%or coincidentally with the GRB. Therefore, the model 
%can be easily tested observationally. If the bursts are 
%observed, we can not only confirm the existence of axion but also detect its
%mass.

Let us first explain briefly axionic boson stars ( ABSs ) 
and how rapidly they dissipate\cite{iwa1} their energies 
in strongly magnetized conducting media just as neutron stars.
The ABS is a coherent object of the real scalar field $a(x)$ describing axion.
It is a gravitationally bound state. 
It is represented by a solution\cite{real,iwa}
of the equation of the axion field coupled with gravity.       
Therefore, it is similar to a soliton although
its stability is not guaranteed topologically. The stability 
has been checked only numerically\cite{real} as
starting with a random configuration 
of the axion field a configuration of the solution appears gradually 
and it seems to be stable. 

An approximate form of the solution\cite{iwa1} is given by

\begin{equation}
\label{a}
a(x)=f_{PQ}a_0\sin(m_at)\exp(-r/R_a)\,, 
\end{equation}
where $t$ ( $r$ ) is time ( radial ) coordinate and 
$f_{PQ}$ is the decay constant of the axion. 
The value of $f_{PQ}$ is constrained\cite{text} from cosmological 
and astrophysical considerations\cite{text,kim} such that 
$10^{10}$GeV $< f_{PQ} <$ $10^{13}$GeV ( the axion mass $m_a$ is 
given in terms of $f_{PQ}$ such that $m_a\sim 10^7\mbox{GeV}/f_{PQ}$ ).
It is allowed in some string models that $f_{PQ}$ ( $m_a$ )
takes a larger ( smaller ) value such as $10^{16}$GeV ( $10^{-9}$eV ).
$R_a$ represents the radius of 
an ABS which has been obtained\cite{iwa1} numerically 
in terms of the ABS mass $M_a$,

\begin{equation}
\label{R}
R_a=6.4\,\frac{m_{pl}^2}{m_a^2M_a}\,,
\end{equation}
Similarly, the amplitude $a_0$ in eq(\ref{a}) is represented by

\begin{equation}
\label{a_0}
a_0=1.73\times 10^6 \frac{(10\mbox{cm})^2}{R_a^2}\,
\frac{10^{-5}\mbox{eV}}{m_a}\,.
\end{equation}
Therefore, 
we find that the solution is parameterized by one free parameter,
either one of the mass $M_a$ or the radius $R_a$ of an axionic boson star.
It is also important to note that the solution is not static but
oscillating with the frequency of $m_a/2\pi$. It has been demonstrated 
that there is no static 
regular solution of the real scalar massless field coupled with gravity.
This may be very general
property even in the real scalar field with mass. On the other hand, 
static solutions\cite{re}
exist in the case of the complex scalar field.

The ABS mass is determined by physical conditions under which the ABS
has been formed; how large cloud of axions are cooled gravitationally
to form the ABS, etc.. The situation is similar to other stars such as
neutron stars or white dwarfs. A typical mass scale in these
cases is the critical mass\cite{star}; 
stars with masses larger than the critical 
mass collapse gravitationally into more compact ones or black holes. 
In the case of the ABS, there also exists a critical mass $M_c$ 
which is given by\cite{real}

\begin{equation}
M_c\simeq 10^{-5}M_{\odot}\frac{10^{-5}\mbox{eV}}{m_a}\,,
\end{equation} 
where $M_{\odot}$ represents the solar mass.
Therefore, we adopt this critical mass $M_c$ 
as a typical mass scale of an ABS present in 
our Universe. Then, the ABS radius $R_a$ for this critical mass
is given such that $R_a\simeq 16\,m_5^{-1}\,\mbox{cm}$. 
This assumption is crucial for our mechanism of generating GRBs.
An ABS with a much smaller mass can not generate GRBs with such a large amount 
of the energy $\sim 10^{50}\,\mbox{erg}$ as observed.

We should mention that the ABS with the mass $M_c$ is not unstable against
the stimulated emission, $a\rightarrow \gamma\,\gamma$, 
argued by Tkachev\cite{decay}.
The emission can occur in an axion medium where both processes of 
$a\rightarrow \gamma\,\gamma$ and $\gamma\,\gamma\rightarrow a$ arise 
efficiently. The radius of the ABS, however is 
nearly equal to the Compton wave length 
of the axion and so the inverse process, $\gamma\,\gamma\rightarrow a$, 
can not occur efficiently.

Now let us explain how rapidly an ABS dissipates its energy
in a neutron star. The essential point is that
an axion is converted into a photon in the external magnetic field $\vec{B}$.
Namely, due to the interaction between the axion and the electromagnetic field
described by
$ L_{a\gamma\gamma}=c\alpha a\vec{E}\cdot\vec{B}/f_{PQ}\pi$, 
the oscillating electric field $\vec{E_a}$ is induced when the ABS is 
exposed to the magnetic field such that 
$\vec{E_a}=-c\,\alpha \,a(x)\,\vec{B}/f_{PQ}\pi$ with $\alpha\simeq 1/137$, 
where the value of $c$ depends on the axion models\cite{DFSZ,hadron,kim} and
typically it is of the order of unity. This can be easily understood by
considering Gauss law modified by $L_{a\gamma\gamma}$\cite{Si,iwa1}.  
Therefore, an ABS induces the oscillating electric current $J_a$ 
in a neutron star.
The current is strong especially in the outer crust of a neutron star
where the density of electrons
is so high that the electric conductivity $\sigma$ is very large, e.g.
$\sigma=10^{26}$/s\cite{con}. The energy of this current is dissipated 
with a rate proportional to $J_a E_a=\sigma E_a^2$. 
Therefore, an ABS dissipates rapidly its energy 
in the magnetized conducting medium owing to  
the large electric conductivity and the strong magnetic field 
of a neutron star.
%Since both of the magnetic field and the electric conductivity
%in a neutron star is very large, the energy dissipation is so rapid that
%a fireball may be generated.

Numerically, the electric field strength and  
the energy dissipation rate $W$ are given by 

\begin{eqnarray}
\label{e}
E_a &\sim& 10^{17}\,\,\mbox{eV}\,\,\mbox{cm}^{-1}\,\,B_{10}\,m_5\, 
\quad \mbox{and}\\
W&=&\int_{ABS}{\sigma E_a^2}d^3x=
4c^2\times 10^{57}\mbox{erg/s}\,\frac{\sigma}{10^{26}/s}\,
\frac{M_a}{10^{-5}M_{\odot}}\,\frac{B^2}{(10^{10}\mbox{G})^2} \\
&\simeq& 4\times 10^{57}\mbox{erg/s}\,\frac{\sigma}{10^{26}/s}\,
\frac{B^2}{(10^{10}\mbox{G})^2}\,m_5^{-1}\,,
\end{eqnarray}
respectively, with $B_{10}=B/10^{10}\mbox{G}$ and $m_5=m_a/10^{-5}\mbox{eV}$.
We have used the solution in eq(\ref{a}) for the critical mass. 
Here, the integration has been performed over the volume of an ABS which 
is located in the outer crust of a neutron star. Note that 
$R_a\simeq 16\,m_5^{-1}\,\mbox{cm}\ll 10^4$ cm. 
Thus it can be inside of the crust whose depth is 
about several hundred meters\cite{star}.

The conductivity quoted above is a theoretical 
expectation in the deep side of the outer crust. 
Therefore, it is larger than a value averaged over the outer crust.
Although the real values may vary depending on neutron stars, 
we may use the value or smaller one
such as $10^{22}$/s, which is expected 
around the envelope of the neutron star. 
In both cases, the rate is very large:
When an ABS with $M_a=10^{-5}M_{\odot}/m_5\sim 10^{49}/m_5\,\mbox{erg}$ 
is located in the crust, it evapolates within $10^{-8}$ seconds 
or $10^{-4}$ seconds 
for $\sigma=10^{26}$/s or $\sigma=10^{22}$/s, respectively. 
Similarly, neutron 
stars may possess magnetic fields 
with various strengths. Depending on the strengths the rate of the dissipation
may change. In any case,   
we find that the electric field strength is huge and that 
the energy dissipation proceeds very rapidly.

Here, we should comment that the energies released in the collisions
are at most the mass of an ABS, $M_a\sim 10^{49}\,m_5^{-1}\,\mbox{erg}$.
Since the lower bound of the axion mass 
is at least $10^{-6}$ eV in the conventinal axion models, 
the energies released are about $10^{50}\,\mbox{erg}$.
It seems that the values are not enough 
for explaning the energies of GRBs observed:
in some events they reach as much as $10^{53}\,\mbox{erg}$. 
However, the energies estimated 
are ones obtained under the assumption of spherical 
explosion of a GRB. As has been pointed out, if the energies of 
GRBs are released in jets with small solid angle  
$\Omega_{GRB}$ such as $10^{-4}$,
the energies are about $10^{49}\,\mbox{erg}$. 
As we discuss below,
such jets can be formed in our model. Thus, the amount of the energies 
$\sim 10^{49}\,m_5^{-1}\,\mbox{erg}$ may be consistent with 
the observation. It is interesting to note that 
a small axion mass such as $10^{-9}\,\mbox{eV}$ is allowed
in a string model\cite{string}. In such case, a large amount of energies 
$\sim 10^{53}\,\mbox{erg}$ can be released.

We wish to see more closely how the energy of an ABS is dissipated
and to estimate how large amount of the energy each particles can obtain 
in the dissipation. Then, we will argue that jets are formed with large 
Lorentz factors and and with small solid angles.
Suppose that the relative velocity $v$ between an ABS and a neutron star
is equal to $10^{-3}\times \mbox{light velocity}$. Furthermore, assume 
that the ABS is trapped to the neutron star 
when it approaches the neutron star
within a distance where the potential energy of ABS around the neutron star is 
larger than the kinetic energy of the ABS. After being
trapped, it is assumed to fall on to the neutron star in a much shorter period
than that in which an ABS is trapped to a neutron star 
after being created.
Then, we think that 
the ABS collides with the neutron star 
with the typical velocity $v$ assumed above.
When the dissipation rate is not large,
it may pass through the neutron star 
without dissipating its whole energy since it only takes 
$5\times 10^4\mbox{cm}/v\simeq 0.0016$ s to pass the outercrust with
its depth assumed to be $5\times 10^4\mbox{cm}$.
Within the period, the ABS dissipates its energy of 
$0.0016\,\mbox{s}\times W\simeq 6\times 10^{50}\mbox{erg}\,
\sigma_{22}\,B_{10}^2\,m_5^{-1}$ ( $\sigma_{22}=\sigma/(10^{22}$/s) ). 
Thus, we understand that if  
$\sigma_{22}\,B_{10}^2 <10^{-1}$,  the ABS mass 
$M_a\sim 10^{49}/m_5\,\,\mbox{erg}$, is not completely 
dissipated in the period. 
In this case, the ABS goes through inside of neutron star and again enters 
the outer crust from the deep inside of the neutron star. 
If the ABS does not lose 
the remaining energy in the region, it can go out of the neutron star.
Then, it goes back to the neutron star and again collides with the latter.
This process continues 
until the whole energy of the ABS is dissipated completely.
The total number of the collisions depends on 
the parameters, $\sigma_{22}\,B_{10}^2$
and the relative velocity $v$. 
A variation of these parameters in the collisions 
makes time-dependent properties of the GRB complex.  
Here, we have assumed implicitly that the energy dissipated is 
negligible in a region below the outer crust. This is because 
electron density or charge density in the region, 
e.g. the region of neutron liquid, is so much small that the induced 
current $\sigma E_a$ is also small.

The energy of a ABS dissipated in the collision is carried away by particles 
composing the outer crust, each of which may obtain much high energy from
the ABS.  
Let us roughly estimate such energy. 
The particles are  
heavy nuclei, protons or electrons 
composing the outer crust.
These particles are 
accelerated by the electric field given in eq(\ref{e})
until they collide with other particles and lose
their obtained energies.
The mean free path $L$ is given by $L=1/nS$ for the number density $n$
of the particles and for the cross section $S$ of the particles. 
Obviously, the mean free path of electrons is largest among the particles and
their cross section $S\simeq \alpha^2/m_e^2$ is smallest ( 
where $m_e$ denotes the electron mass ). 
Hence, the electrons can obtain much higher energies
than those did by the other particles. Their energies in turn can be 
transformed to the other particles with collisions. Thus, the energy 
$\Delta E$
obtained by electrons may be regarded as the typical energy obtained by 
the particles,

\begin{eqnarray}
\Delta E&=&e\,E_a\,L=e\,E_a/nS=e\,E_a\,m_e^2/n\alpha^2 \\
        &\sim& 10\,\mbox{GeV}\,B_{10}\,m_5/n_{34}\,,
\end{eqnarray}
where $n_{34}=n/10^{34}$ and the number density $n_{34}=1$ 
roughly corresponds to the mass density $10^{10}$ g of the outer crust. 
This is an underestimation of the energy $\Delta E$ because the cross section
of electrons accelerated by the field is much smaller than that 
( $\sim \alpha^2/m_e^2$ ) used in the estimation. It is given by 
$\sim \alpha^2/(\Delta E)^2$ if the energy $\Delta E$ obtained by electron is 
much higher than $m_e$. 
Therefore, we can expect that when the particles are ejected, 
actual value of $\Delta E$ is 
much larger than $10\,\mbox{GeV}\,B_{10}\,m_5/n_{34}$. 
For example, as the particles obtain higher energies, 
the mean free path becomes longer
since the cross section becomes smaller. Furthermore, 
the longer mean free
path leads to much higher energy obtained by the particles. 
Consequently, it turns out that  
the mean free path of electron becomes infinite 
in the case of a small density ( $n <10^{34}$ ). 
In such case,
the particles are accelerated only within a period $2\pi/m_a$
of oscillation of the electric field and they obtain very high energy of 
$\Delta E=e\,E_a\,2\pi/m_a\,
\times \mbox{light velocity}\sim 10^{18}\,\mbox{eV} B_{10}$. 
Although we have neglected the effect of magnetic fields in the estimation,
we can see that the particles ejected carry much high energies.
Therefore, we can expect that the ejected particles may have much large 
Lorentz factors such as $10^3$ which are needed for fireballs leading to
GRBs observed. Actually, the ejected particles come from 
the outercrust of a neutron star. Mass fraction of the outercrust is 
less than $10^{-6}$ of the total mass $\sim M_{\odot}$ 
of the neutron star. Therefore, They have large Lorentz factors.

We should stress that the particles are accelerated into a direction
parallel to the magnetic field $\vec{B}$ since 
the electric field $\vec{E_a}$ is parallel to $\vec{B}$.
Consequently, they form 
jets with a very small solid angle 
$\Omega \sim (1\,\mbox{MeV}/\Delta E)^2\ll 10^{-8}$
where particles are assumed to have a kinetic energy 
of the order of $1\mbox{MeV}$
in the neutron star. Although this is too naive estimation, the result 
indicates that the jets may have very small solid angle.  
This is a clear prediction of our model for
a mechanism generating GRBs. 
Actual process is, however, more complicated than this naive one. 
When the electric field is pointed at 
the outside of the neutron star,
electrons are accelerated to the inward but 
some of them are scattered outward. On the other hand, when
the electric field is pointed at the inside, electrons are 
accelerated to the outward and fewer are scattered inward 
than in the former case.
Through these processes real jets may be generated with small solid angles.

Taking account of these facts, we can understand qualitatively 
some of time-dependent 
features of GRBs. First, we find quiet intervals
between emission episodes in some GRBs. This is caused by 
the fact that the ABS passes through the
neutron star in the first collision and then goes back, 
colliding with the latter again. The emission
does not occur until the ABS collides with the neutron star again after
passing it.
Second, we find 
a single pulse without quiet intervals in some GRBs. This is due to 
the fact that the ABS can not pass the neutron star 
or dissipate the whole energy
in the first collision. Third, we find bimodality of pulses in some GBRs.
These are pulses with a short duration and pulses with a long duration.
This is caused by the fact that when the ABS passes 
the neutron star with 
a relatively high velocity 
in the first collision, the sharp pulse with a short duration may appear.
Furthermore, in the second collision the ABS collides 
with a smaller velocity than  
in the first collision, which gives rise to   
a pulse with a long duration: 
the smaller velocity is expected in the second collision,
since the kinetic energy of the ABS may be dissipated 
in the first collision, in general.
These interpretations of the time-dependent properties of GRBs are 
natural consequences of our mechanism.

Now, we wish to point out the important prediction of intriguing phenomena 
expected in our mechanism.
As we have mentioned before, the current induced by an ABS in a neutron star
is oscillating so that electromagnetic radiations are emitted by the current.
Let us calculate the luminosity of such radiations.
The electromagnetic potential $A_i$ of the radiations generated 
by the current $J_a$ is given by

\begin{eqnarray}
A_i&=&\frac{1}{R_0}\int J_a(t-R_0+\vec{x}\cdot\vec{n})\,d^3x\\
&=&\frac{c\alpha\sigma a_0B_i}{\pi R_0}\int
\sin m_a(t-R_0+\vec{x}\cdot\vec{n})\,\exp(-r/R_a)d^3x\\
&=&\frac{4c\alpha\sigma a_0B_i}{R_0}\,\frac{R_a^3}{(m_a^2R_a^2+1)^2}\,
\sin m_a(t-R_0)\,,
\end{eqnarray}
where $R_0$ ( $\vec{n}$ ) denotes a distance ( direction )
between ( from ) the neutron star and ( to ) the earth.    
We have integrated the current $J_a$  over the ABS, which is involved in the 
envelope or the crust of the neutron star. Thus, 
the luminosity $L$ of the radiations with the frequency, 
$m_a/2\pi=2.4\times 10^9\,m_5\,\mbox{Hz}$ is given by

\begin{eqnarray}
L&=&4 c^2 \alpha^2 \sigma^2 a_0^2 m_a^2 B^2 \frac{R_a^6}{(m_a^2R_a^2+1)^4}\\
&\sim&10^{47}\mbox{erg/s}\,B_{10}^2\,\sigma_{15}^2/m_5^2\,,
\end{eqnarray}
where we have taken a small value of $\sigma$ as a reference since 
the radiations capable for going out of the neutron star only appear near 
the envelope. If the radiations appear in the deep side of the crust,
they are absorbed into the medium and can not go out of the neutron star
so that they are unobservable.
It is difficult to estimate the precise value of the luminosity 
because we do not 
have enough knowledge about the conductivity $\sigma$ or optical depth
of the radiations with the wave length of $2\pi/m_a$.
In particular, the optical depth heavily depends 
on the composition, temperature or 
density of the envelope. 
Hence, the above estimation 
should be regarded as a rough estimation of the order of magnitude. However, 
it indicates that the luminosity of the radiations may be 
large enough to be detectable. 
Therefore, we can expect that the burst of the radiations
coincides with a GRB. These radiations must have 
the wave length given by $2\pi/m_a $ ( $=12.6\,\,\mbox{cm}/m_5$ ).
Thus, with observation of them we can not only test our model for GRBs 
but also determine the axion mass $m_a$.

Let us comment that the burst appears only when an ABS passes 
the envelope of the 
neutron star. Therefore, the duration of the burst must be much less than 
$10^{-3}$ sec. Furthermore, 
it arrives at the earth just before or coincidentally with 
the GRB. This is because the burst appears just before the appearance  
of jets leading to GRB; the jets are produced in the crust while 
the burst of the monochromatic radiations is produced in the envelope. 
Most of the energies associated with 
these phenomena may be carried by the jets and probably only a fraction of them
is carried away by the burst of the monochromatic radiations.

Finally, we show that the rate of collisions between a neutron star
and an ABS is consistent with observations. To this end, 
we suppose that the total number of the neutron stars in a galaxy is 
$\sim 10^9$.  
Let us assume that the halo of the galaxy
is composed mainly of ABSs whose typical velocity $v$ is supposed to be 
$3\times 10^7$ cm/s\cite{text}. 
Since the local density of the halo in our galaxy is given by 
$0.5\times 10^{-24}$g $\mbox{cm}^{-3}$\cite{text},  
the rate $R_c$ of collisions per year and per a galaxy 
is calculated as follows:

\begin{equation}
R_c=n_a\times N_{ns}\times S_c\,v\times 1\,\mbox{year}
\simeq 10^{-6}m_5\,\,
\mbox{per year}
\end{equation}
with $n_a=0.5\times 10^{-24}\mbox{g cm}^{-3}/M_a$ being 
the number density of ABSs
in the galaxy and with $N_{ns}\sim 10^9$ being the number of the neutron stars.
The cross section $S_c$ for such collision can be estimated 
in the following:
an ABS is trapped to a neutron star 
when they approach within a distance $L_c$
and then it collides with the latter. The distance $L_c$ is 
determined such that the kinetic energy 
$v^2M_a/2$ of ABS ( where the relative velocity is assumed to be 
$v$ ) is equal to  
the potential energy $GM_aM_{\odot}/L_c$ of the ABS 
in the gravitational field of
the neutron star, i.e. $L_c\sim 10^{11}$ cm.
Thus, $S_c=\pi L_c^2$. 
Since $S_c$ ( $\propto v^{-4}$ ) is strongly dependent 
on the assumed velocity $v$,
the value of the rate $R_c$ may varies from $10^{-3}$ to $10^{-9}$
corresponding to the variation of 
$v$, $3\times 10^{6}$ cm/s to $3\times 10^{8}$ cm/s.
The observed rate $R_{obs}$ is of the order of 
$\sim 10^{-8}$ per year and per galaxy. 
Thus, taking account of the fact that 
GRBs are emitted into a cone with the solid angle $\Omega_{GRB}$, 
we obtain a relation, 

\begin{equation}
R_{obs}\simeq 10^{-8}=(\Omega_{GRB}/4\pi)\,R_c\,.
\end{equation} 
This is satisfied even by the value of $\Omega_{GRB}=10^{-4}$ if 
we assume a small velocity of $v=10^{-4}\times \mbox{light velocity}$ 
in the collision, i.e. $R_c\simeq 10^{-3}$. 
However, it is very improbable for jets to 
have a solid angle such as $\Omega_{GRB}=10^{-6}$.

In summary, we have analyzed a model for an engine of GRBs 
proposed by the present author and 
have shown that the collision between an axionic boson
star and a neutron star generates a jet of baryons and leptons, whose 
Lorentz factors must be larger than $10^2$. 
The collision may occur several times  
when the axionic boson star does not dissipate its whole energy 
in the first collision.
These multiple collisions are taken as the origin of 
time-dependent complex properties of 
GRBs. We have also estimated the rate of collisions and found that 
although the rate is quite sensitive to the relative velocity,
it is consistent with the observations.  
A most important prediction of our model is 
the emission of the radio waves with 
the frequency given by the axion mass. The emission appears 
coincidentally with GRBs. Therefore, we can test the validity 
of our model by detecting these radio waves.
 
\vskip .7cm

The author wish to express his thank to Prof.Terazawa
for careful reading the manuscript and to all staffs in Theory Group,
Tanashi Branch, High Energy Accelerator Research Organization 
for the warm hospitality extended to him.
This work is supported by the Grant-in-Aid for Scientific Research
from the Ministry of Education, Science and Culture of Japan No.10640284

%%%%%%%%%%%%%%%%%%%%%%


\begin{thebibliography}{99}
\bibitem{model}P. M\'esz\'aros, astro-ph/9904038, and references cited therein.
\bibitem{text}For a review, see, for example, 
E.W. Kolb and M.S. Turner, The Early Universe, Addison-Wesley, New York,
(1990).
\bibitem{PQ}R.D. Peccei and H.R. Quinn, Phys. Rev. Lett. 38, 1440 (1977),\\
S. Weinberg, Phys. Rev. Lett. 40, 223 (1978),\\
F. Wilczeck, Phys. Rev. Lett. 40, 279 (1978).
\bibitem{kim}J.E. Kim, Phys. Rep. 150, 1 (1987).
\bibitem{iw}A. Iwazaki, Phys. Lett. B406, 304 (1997); 
Phys. Rev. Lett. 79, 2927 (1997). 
\bibitem{s}P. Sikivie, Phys. Rev. Lett. 51, 1415 (1983).
\bibitem{cooling}E. Seidel and  W.M. Suen, Phys. Rev. Lett. 72, 2516 (1994).
\bibitem{kolb}E.W. Kolb and I.I. Tkachev, Phys. Rev. Lett. 71, 3051 (1993);\\
Phys. Rev. D49, 5040 (1994).
\bibitem{iwa}A. Iwazaki, Phys. Lett. B455, 192 (1999).
\bibitem{real}E. Seidel and W.M. Suen, Phys. Rev. Lett. 66, 1659 (1991).
\bibitem{iwa1}A. Iwazaki, Phys. Lett. B451, 123 (1999), \\
Phys. Rev. D60 025001 (1999).
\bibitem{complex}B. Stern, astro-ph/9902203.
\bibitem{re}P. Jetzer, Phys. Rep. 220, 163 (1992),\\ T.D. Lee 
and Y. Pang, Phys. Rep.
221, 251 (1992),\\ A.R. Liddle and M. Madsen, 
Int. J. Mod. Phys. D1, 101 (1992).
\bibitem{star}S.L. Shapiro and S.A. Teukolsky, Black Holes, White Dwarfs, 
and Neutron Stars, A Wiley-Interscience Publication, (1983).
\bibitem{decay}I.I. Tkachev, Phys. Lett. B191, 41 (1987).
\bibitem{DFSZ}A.R. Zhitnitsky, Sov. J. Nucl. Phys. 31, 260 (1980),\\ M.D. Dine,
W. Fischler and M. Srednicki, Phys. Lett. B104, 199 (1981).
\bibitem{hadron}J. E. Kim, Phys. Rev. Lett. 43, 103 (1979),\\ M.A. Shifman,
A.I. Vainshtein and V.I. Zakharov, Nucl. Phys. B166, 493 (1980).
\bibitem{Si}P. Sikivie, Phys. Lett. B137, 353 (1984).
\bibitem{con}D.A. Baiko and D.G. Yakovlev, Astron. Lett. 72, 702 (1995);
astro-ph/9604164.
\bibitem{string}T. Banks and M. Dine, Nucl. Phys. B479, 173 (1996);
Nucl. Phys. B505, 445 (1997).
%\bibitem{ba}M.J. Ree and P. Meszaros, MNRAS, 41P, 258 (1992).
%\bibitem{grb}S. Odewahn, J. Bloom and S. Kulkarni, G.C.N. 201 (1999).
%\bibitem{kolb2}E.W. Kolb and I.I. Tkachev, Phys. Rev. D50, 769 (1994).



\end{thebibliography}
\end{document}